\documentclass[aps, nofootinbib, twocolumn]{revtex4}
\usepackage{graphicx,amsfonts}

\newcommand{\Ref}[1]{(\ref{#1})}

\def \be{\begin{equation}}
\def \ee{\end{equation}}
\def \bes{\begin{eqnarray}}
\def \ees{\end{eqnarray}}

\newcommand{\Cb}{{\rm \bf C}}

\def \sl2{SL(2,\Cb)}

\begin{document}

\title{Generalised group field theories and quantum gravity transition
amplitudes}

\author{{\bf Daniele Oriti}} 
\affiliation{\small Department of Applied Mathematics and Theoretical
  Physics, \\
University of Cambridge, Wilberforce Road, Cambridge CB3 0WA, England,
EU \\ e.mail: d.oriti@damtp.cam.ac.uk}
\date{\today}

\begin{abstract}
We construct a generalised formalism for group field theories, in
which the domain of the field is extended to include additional proper
time variables, as well as their
conjugate mass variables. This formalism allows for different types of
quantum gravity transition amplitudes in perturbative expansion, and
we show how both causal spin foam models and the usual a-causal ones
can be derived from it, within a sum over triangulations of all
topologies. We also highlight the relation of the so-derived causal
transition amplitudes with simplicial gravity actions. 
\end{abstract}
\maketitle

\section{Introduction}
Spin foam (SF) models \cite{review, alex} are a promising recent approach
to quantum gravity, in any dimension, as an algebraic and combinatorial sum-over-histories. They are defined by a sum
over all possible spacetime geometries, encoded
in a 2-complex $\Gamma$ labeled by representations of the Lorentz
group, weighted by a quantum amplitude function of these algebraic
data and in general depending also on the combinatorics of the
underlying 2-complex, with given boundary data. In turn, SF
models are \cite{mikecarlo} Feynman
amplitudes of so-called group field theories (GFT), field theories defined
on appropriate group manifolds, and the sum over spin foams is the
sum over Feynman graphs of the corresponding field theory. The most studied
model in 4-d is the Barrett-Crane (BC) model \cite{BC}:
$$ 
Z(\Gamma)= \sum_{\{J_f\}}\prod_f \Delta_{J_f}\prod_e
A_e(J_{f\mid e})\prod_v A_v(J_{f\,id v})
$$
in which $\Gamma$ is topologically dual to a simplicial complex, and where the sum is over class I representations
$J$ of $SL(2,\mathbb{C})$ or $Spin(4)$, depending on the signature,
assigned to the faces of $\Gamma$, with $\Delta_J$
their Plancherel measure for $SL(2,\mathbb{C}))$ or their dimension
for $Spin(4)$, and the quantum
amplitude is factorised into face $f$, edge $e$
and vertex $v$ contributions. The vertex amplitude is \cite{laurentkirill}:
$$
A_v^{BC}=\prod_{e\mid v}\int_{G}dg_{e\mid v}\prod_{f\mid v}
H^{J_f}(g_{e1,g_{e2}})=\begin{array}{c}\includegraphics[width=1.5cm]{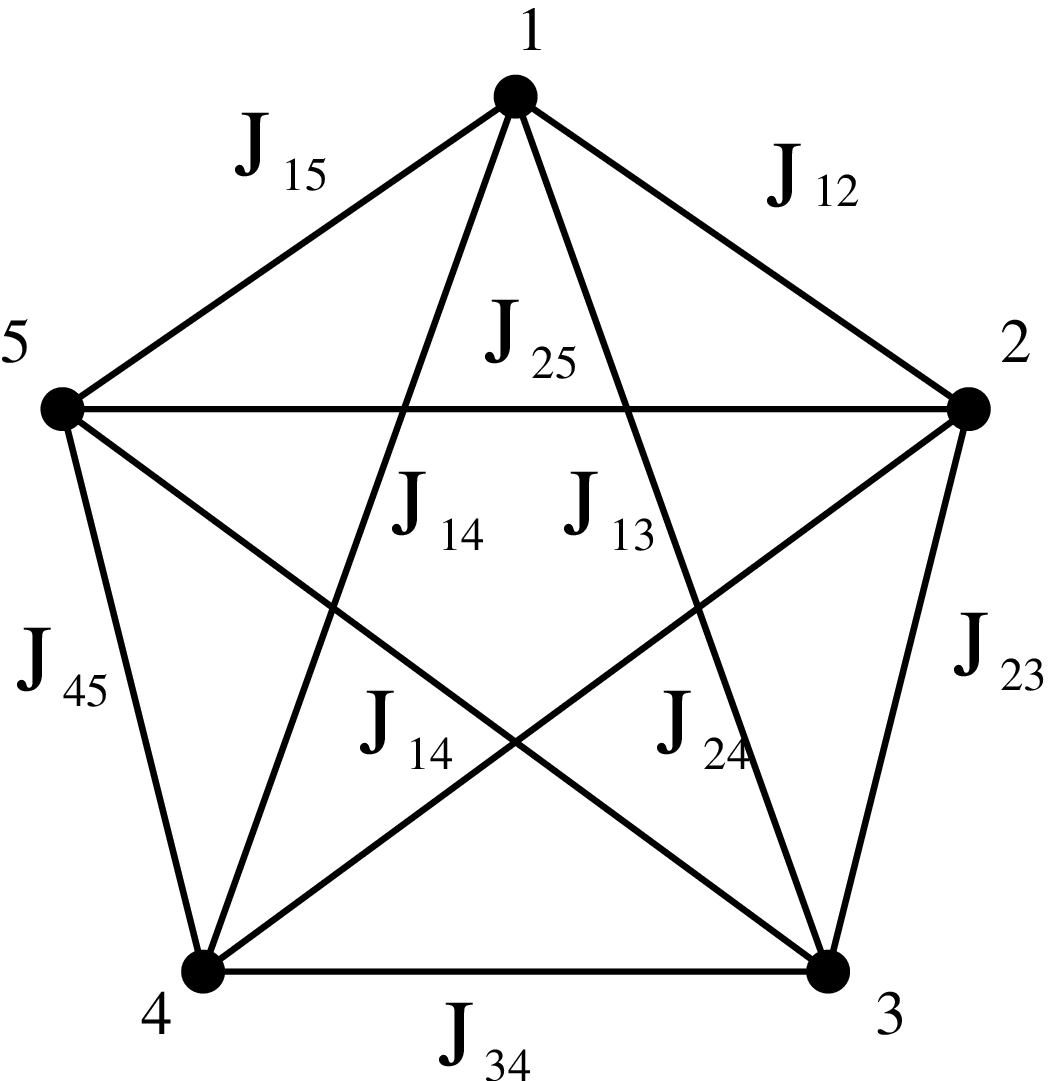}
\end{array}\nonumber,
$$  
where the five edges of $\Gamma$ incident to the vertex $v$ are
represented by a point and the 10 faces incident to the
same vertex, each bounded by 2 of the edges, are represented by a line;
the integral kernel $H^\rho(g_{e1},g_{e2})$ is a zonal spherical
function $D_{00}^\rho(g_{e1}g_{e2}^{-1})$ for the group $G$ \cite{laurentkirill}.   
The various versions of this model, differing for the edge
amplitude $A_e$, can all be obtained
through a GFT \cite{DP-F-K-R,P-R,review}. 
The relevant field is a complex scalar field defined over the product
of 4 copies of $G$, with a global gauge invariance
property: $\phi(g_1,g_2,g_3,g_4)=\int_G dg \phi(g_1g,g_2g,g_3g,g_4g)=\phi(g_i)$,
and symmetric under even permutations of its arguments; the action for
the theory \cite{DP-F-K-R,P-R} is: 
\bes 
\lefteqn{S(\lambda)=\prod_{i=1}^{4}\int
dg_i\,P_h\phi(g_i)P_h\phi(g_i)+\frac{\lambda}{5}\prod_{i\neq
  j=1}^{5}\int_G dg_{ij}} \nonumber \\ &\left\{ P_h\phi(g_{1j})
P_h\phi(g_{2j})P_h\phi(g_{3j})P_h\phi(g_{4j})P_h\phi(g_{5j})\prod\delta(g_{ij},
g_{ji})\right\} \nonumber
\ees
where $P_h\phi(g_i)=\prod_i\int_{SU(2)}dh_i\phi(g_ih_i)$; if the
projections $P_h$ are dropped from the kinetic term, one obtains the
version of \cite{P-R}. $\phi$ represents a
2nd quantized tetrahedron, its 4 arguments its 4
triangles, and the interaction term has indeed the combinatorial
structure of a 4-simplex with 5 tetrahedra glued along triangles. The
perturbative expansion produces a sum over Feynman graphs, 2-complexes dual to 4d triangulations, with amplitudes
given by the BC model:
\be
\mathcal{Z}(\lambda)=\sum_{\Gamma}\frac{\lambda^N}{sym(\Gamma)}Z(\Gamma).   
\ee
In
\cite{causal,feynman} a class of spin foam models, incorporating extra causality restrictions,
and interpreted as the quantum gravity analogue of the Feynman propagator
or causal 2-point function of QFT, has been
constructed \cite{feynman}. In this paper we present a generalised
GFT formalism from which models of the type introduced in \cite{feynman} as well as
the usual ones can be derived, differing in their
causal properties in the sense of \cite{feynman}, confirming their respective
interpretation as causal and a-causal transition amplitudes for
quantum gravity.  Our construction and results apply to any
spacetime dimension and any signature; however we present here only
the 4d case because of more direct physical interest, and work in
Riemannian signature for simplicity of notation, but the translation
of notation and results to the Lorentzian signature is
straightforward. Also, we do not discuss in detail the analytic
continuation needed to define some of the integrals involved in our
contruction, of the very same type as that for Feynman propagators in
QFT. 

\section{Motivation and broader picture}
Our main motivation for looking for new SF models for quantum
gravity is the idea that one should be able to define
more than one type of transition amplitudes for it, just as in QFT (or the relativistic particle). A
related motivation is to incorporate
causality in SFs, given that the difference between
various transition amplitudes in QFT is in their different causal
properties. The idea of doing this in quantum gravity is not new. In \cite{teitelboim}
it is shown that in a phase space path integral for gravity the choice
of the range of integration over the lapse function characterizes the
difference between the a path integral definition of an analogue of
the Hadamard function (infinite range $(-\infty,+\infty)$),
a projection onto solutions of the Hamiltonian constraint,
and of the analogue of the Feynman propagator (half-infinite range
$(0,+\infty)$), a causal transition amplitude \footnote{The first is a solution
of the Hamiltonian constraint operator, while the second
is a Green function for it;
however, both are fully invariant with respect to 4-dimensional
diffeomorphisms (the positive half-infinite range is diffeo-invariant and
isomorphic to the negative one), so that their differences come from the difference between diffeomorphisms and
canonical symmetries.}. Both can be given a Lagrangian path integral
realization\cite{halliwellhartle}: given that opposite signs of the
lapse correspond to opposite signs for the
gravity action
(opposite global orientations), the \lq\lq Feynman propagator\rq\rq
for quantum gravity is obtained by a sum over 4d geometries with
amplitude given by the exponential of the gravity action, and the
\lq\lq Hadamard propagator'' is obtained summing over {\it the same}
set of geometries with amplitude given by the sum of two exponentials
of the gravity action with opposite signs:
$$
Z_F=\int\mathcal{D}g\,e^{iS(g)} \nonumber 
\hspace{0.2cm} vs \hspace{0.2cm} Z_H=\int\mathcal{D}g\left( e^{iS(g)} +
e^{-iS(g)}\right).
$$
As in QFT, the difference is a causal restriction: considering only positive proper times or symmetrizing over them. This is equivalent to summing only over one choice of global spacetime
orientation, even if locally any space orientation is
considered, or symmetrizing over it. The analogy with QFT is even more
clear from the point of view of 3rd quantization of gravity
\cite{giddingsstrominger}, where the gravitational Lagrangian path integral is
indeed a Feynman 2-point function for the field in superspace.

\vspace{0.1cm}

We find a similar situation also in SF models, as it is clear from the asymptotic
analysis of the BC vertex amplitude \cite{asymp}, that
gives indeed the cosine of the Regge action, and by the fact that all known SF models define real amplitudes and are symmetric under change of orientation, because of the use in each dual face of the kernels $H^J(g_1,g_2)$
that are the real part of an orientation dependent complex amplitude \cite{causal, feynman}. A class of SF models that breaks this symmetric
structure by substituting the kernels $H^J$ with appropriate
generalisation of them using a proper time parametrization, and that
reduce to the $H^J$ is the infinite range of integration over it is chosen,
while producing new causal amplitudes when the half-infinite range is
selected, was constructed in \cite{feynman}. The motivation was to realise in spin foam quantum gravity what is only formally
realised in the continuum path integral quantization. The importance of these new
models lies also in that they are given by the exponential of it times a complicated measure) \cite{feynman}, and thus in a direct link to other
approaches to simplicial quantum gravity like quantum Regge calculus and
causal dynamical triangulations\cite{DT}.

\vspace{0.1cm}
   
What is the need
for a GFT derivation of the causal models and for a GFT implementation
of the above ideas?  First of all, one cannot really speak of a causal {\it model}, without a clear prescription for all its amplitudes, thus until some sort of derivation is
performed. In
fact, \cite{feynman} gave a proposal for the
vertex amplitudes of causal models, but not a complete definition
of them. Also, the GFT formalism allows to get rid of the
dependence on a fixed 2-complex and to realise the sum over
2-complexes in a natural
way. There is a more fundamental motivation, however: the idea that
the group fied theory represents the truly fundamental definition of a
quantum gravity theory based on spin foam structures. Let us clarify
what is at stake here, from this perspective, and thus what is the
importance of our results. GFTs can be seen as
providing a simplicial 3rd quantization of gravity
\cite{laurentgft,iogft}, purely algebraic and
combinatorial, in which both geometry and topology are dynamical. From this point
of view, the idea of
different types, causal and a-causal, of transition amplitudes with
the same role as in ordinary QFT is indeed more than
an analogy, and one expects to realise here rigorously what was
formally proposed in continuum 3rd quantization theories and in the
path integral formalism. There is more. Group field
theories have all the ingredients that enter other approaches to
quantum gravity: boundary states given by spin networks, as in loop
quantum gravity, a simplicial description of spacetime and a sum over
geometric data, as in Quantum Regge calculus, a sum over
triangulations dual to 2-complexes, as in dynamical triangulations, a
sum over topologies like in matrix models, of which GFTs are indeed
higher-dimensional analogues, an ordering of fundamental events
(vertices of Feynman diagrams), given by the orientation of the
2-complex, which has similarities to that defining causal
sets. Therefore one can envisage GFTs as a {\it general
framework for non-perturbative quantum gravity}. This is at present not
much more than a dream, but clearly a direct connection to the Regge action,
obtained through a GFT that has the exponential of it as
quantum amplitude, and an explicit dependence on orientation data,
would be an important step in extablishing links with the other
approaches. A GFT that produces causal spin foam amplitudes, which seem
to have all these properties, is therefore crucial.  
     
\section{Causal SF vertex amplitudes}
Let us review the construction of \cite{feynman}, and
highlight the ingredients we need in the generalised GFT
formalism. Given a scalar particle with mass $m^2$, living in the
homogeneous space $G=Spin(4)/SU(2)$, one defines a propagator by:$G(g_1,g_2,m^2)=\int ds K(g_1,g_2,s) e^{i m^2 s}$, where $K$ is the evolution kernel in proper time $s$. According
to the range of integration chosen, one obtains either
$G_H(g_1,g_2,m^2)\propto H^J(g_1,g_2)= \frac{i
  e^{i(2J+1)\theta}}{\sin\theta} + \frac{-i
  e^{-i(2J+1)\theta}}{\sin\theta}$, i.e. the Hadamard propagator, with momentum square $C(J)=2J(2J+2)=-m^2$ where $C(J)$
is the Casimir of $Spin(4)$ in the representation $J$, or the Feynman propagator
$G_F(g_1,g_2,m^2)\propto \frac{i
  e^{i\sqrt{1-m^2}\theta}}{\sin\theta}$. The idea is therefore of substituting
$G(g_1,g_2,m^2)$ for $H^J(g_1,g_2)$ in the BC vertex
amplitude to obtain a generalised model and from this (integrating $s$
over an half-infinite range only) the causal models\footnote{In
  \cite{feynman} the square mass $m^2$ was identified with the
  momentum square $C(J)$ also in the Feynman propagator, while this is
  correct strictly speaking only on-shell, thus only for the Hadamard
  propagator.}. Therefore the resulting model will treat
the masses $m^2$ for the \lq\lq particles\rq\rq associated to the
faces of the dual 2-complex as {\it variables}, conjugate to the
proper time variable $s$\cite{Feynman,hostler}. The new kernels $G(g_1,g_2,m^2)$ can be expanded in modes using harmonic analysis, giving:
\bes
G_H(g_1,g_2,m^2)= \int_{-\infty}^{+\infty}ds\sum_J \Delta_J
D^J_{00}(g_1g_2^{-1})e^{-i\left( C(J)+m^2\right) s} \nonumber \\
\longrightarrow \delta\left( C(J)+m^2\right)\,D^J_{00}(g_1g_2^{-1})&&
\nonumber \ees
so that in a model with variable mass, thus involving
an integral over $m^2$, we get back the usual face contribution to the BC amplitude, and
\bes
G_F(g_1,g_2,m^2)=\int_{0}^{+\infty}\sum_J \Delta_J
D^J_{00}(g_1g_2^{-1})e^{-i\left( C(J)+m^2\right) s} \nonumber \\
\longrightarrow \frac{1}{C(J)+m^2-i\epsilon}\,D^J_{00}(g_1g_2^{-1}),
\nonumber
\ees
with $\frac{1}{C(J)+m^2-i\epsilon}$ being indeed the Feynman
propagator in momentum
space.
The vertex amplitude we seek to reproduce from a
generalised GFT is thus of the type:
\bes
A_v=\prod_{e\mid v}\int_{G}dg_{e\mid v}\prod_{f\mid v}
\frac{1}{C(J_f)+m_f^2-i\epsilon}\,D^{J_f}_{00}(g_{e_1}g_{e_2}^{-1})=
\nonumber \\ =\prod_{f\mid v}
\frac{1}{C(J_f)+m_f^2-i\epsilon}\begin{array}{c}\includegraphics[width=1.5cm]{4simp.eps}
\end{array}&&\nonumber,
\ees
within a model involving both a sum over representations $J$
associated to the faces of the 2-complex and an integral  over the
real line for the $m^2$ variables also associated to the faces of the
2-complex. The new causal vertex amplitudes are
given then by a product of Feynman propagators in momentum space with
variable mass, one for each face of the 2-complex,
intertwined by the usual BC vertex amplitude. This is a full momentum
space representation, while we expect a configuration space
representation GFT to involve on equal footing group variables and
proper time variables. 
On top of these, additional data encoding the orientation of the
2-complex will affect non-trivially the amplitudes \cite{feynman};
these can enter the amplitudes in various ways, but it will
be uniquely determined from the generalised GFT formalism.
 
\section{Generalised GFT formalism} 
Consider a complex field: $\phi(g_1,s_1; g_2,s_2;g_3,s_3;g_4,s_4) :
(Spin(4)\times\mathbb{R})^{\otimes 4} \rightarrow \mathbb{C}$, i.e. a
simple extension of the usual field of GFTs to include
a dependence on a proper time variable for each of its arguments\cite{Feynman,hostler}. The
field is required to be invariant under even permutation of its
arguments, while odd
permutations correspond to complex conjugation \cite{DP-P}; this ensures that only orientable
2-complexes are obtained in the perturbative expansion of the theory;
we label by a parameter $\alpha=\pm 1$ the field to represent both the
field and its complex conjugate: $\phi^\alpha(g_i,s_i)/
\phi^+(g_i,s_i)=\phi(g_i,s_i), \phi^-(g_i,s_i)=\phi^*(g_i,s_i)$ (so
that $\alpha$ labels the order of the permutation of the arguments of
the field). The field is required to be invariant under the natural
extension of the usual GFT gauge symmetry, i.e. we now require
invariance under $G\times\mathbb{R}$,
imposed through a projector $P_g$ defined as: $P_g\phi^\alpha(g_1,s_1;
g_2,s_2;g_3,s_3;g_4,s_4)=\int_G
dg\int_\mathbb{R}ds\,\phi^\alpha(g_1g,s_1+s;
g_2g,s_2+s;g_3g,s_3+s;g_4g,s_4+s)$. 
The usual projector $P_h$ is also imposed in the action.
The field is expanded in modes (momentum space) as: 
\bes
\phi^\alpha(g_i,s_i)=\sum_{J_i,\Lambda,k_i}\int_{\mathbb{R}}dm^2_i\left(\phi^{J_i\Lambda}_{k_i}(m_i^2)\right)^\alpha\;\;\;\;\;\;\;\;\;\;\;\;
\nonumber \\ \prod_i
D^{J_i}_{k_il_i}(g_i)C^{J_1..J_4\Lambda}_{l_1..l_4}
\prod_{i}e^{-i m^2_i s_i}\,\delta\left(\sum_i m^2_i\right)\;\;&&
\ees
where we
have redefined the modes of the field contracting them with a
$Spin(4)$ intertwiner $C^{J1..J_4\Lambda}_{k_1..k_4}$ among the
representations $J_1,...,J_4$, with $\Lambda$ labelling a basis in the
space of such intertwiners \cite{P-R}. Because of the
gauge invariance we have imposed,
the field is a combination of $Spin(4)$ and $\mathbb{R}$ intertwiners
(delta functions constraining the sum of the 4 representations $m^2_i$
to be zero). The geometric interpretation of the field is again of a
2nd quantized tetrahedron, now parametrised
by 4 extra proper time variables (treated as independent variables)
one for each of its 4 triangles, with the requirement that the
triangles close to form indeed a tetrahedron (gauge invariance under
$G$) and that the proper time evolution of each of them is defined up
to a global shift of the origin (gauge invariance under
$\mathbb{R}$).
The action is given by:
\bes
\lefteqn{S_{gen}(\{\lambda\})=\sum_{\mu,\tilde{\mu},\alpha,\beta}\frac{1}{4}\prod_{i=1}^{4}\int
dg_i\,\int_\mathbb{R}ds_i  } \nonumber \\ &&\left\{
\phi^{\tilde{\mu}\beta}(g_i,s_i)\left[\prod_i\left(
  -i\mu\alpha\partial_{s_i}+\nabla_{i}\right)\right]\phi^{\mu\alpha}(g_i,s_i)\right\} + \nonumber \\ &+&\sum_{\mu}\sum_{\alpha_i,\epsilon_{ij}}\frac{\lambda_{\{\alpha_i\}}}{5}\prod_{i\neq
  j=1}^{5}\int_G dg_{ij}\int_\mathbb{R}ds_{ij}\nonumber \\ && \left\{
P_h\phi^{\mu\alpha_1}(g_{1j},s_{1j}) 
P_h\phi^{\mu\alpha_2}(g_{2j},s_{2j})P_h\phi^{\mu\alpha_3}(g_{3j},s_{3j}) \right.\nonumber  \\ &&\left.P_h\phi^{\mu\alpha_4}(g_{4j},s_{4j})P_h\phi^{\mu\alpha_5}(g_{5j},s_{5j})
\right.\nonumber \\
&&\left.\prod\,\theta(\epsilon_{ij}(s_{ij}-s_{ji}))K\left(g_{ij},
g_{ji};\mu\epsilon_{ij}(s_{ij}-s_{ji})\right)\right\}\;\;\;\;
\label{gen} \ees
with $\lambda_{\{\alpha_i\}}=\lambda^*_{\{-\alpha_i\}}$, in order to
have an action that is real, where $\nabla_i$ is the Laplacian on the
group manifold $G_i$, and we have assumed an ordering of the fields in the vertex term,
corresponding to an ordering of the tetrahedra in the corresponding
4-simplex. We have also introduced an extra set of data $\mu=\pm 1$ and $\epsilon_{ij}=\pm 1$, being used in the same way as the $\alpha$ to indicate complex conjugation. We interpret the $\alpha$ labelling the
field $\phi^{\alpha}$ as characterizing the orientation of the
tetrahedron corresponding to it with respect to the 4-simplex corresponding to the potential term in the
action, whose orientation is labelled by the variable $\mu_v$,
while the $\epsilon_{ij}$ characterizes the orientation
of the triangle dual to the face bounded by the dual edges
labeled $i$ and $j$ \cite{feynman}. Consistency of the above
interpretation requires then these data to be restricted in the same
way in which the orientation data for a 2-complex are restricted
\cite{causal,feynman}: $\epsilon_{ij}=\alpha^v_i\alpha^v_j$ in the vertex
term and $\mu_{v_1}\alpha^{v_1}=-\tilde{\mu}\beta$ in the kinetic
term. With this choice of orientation variables, the action is real as
it should. This action, therefore, generates oriented
2-complexes as Feynman graphs, with orientation given by the variables
$\mu$ and $\alpha$, that are summed over in the perturbative
expansion, with amplitudes depending non-trivially on this orientation, as we are now going to show. From the kinetic term one deduces the propagator 
in momentum space:
\bes
\mathcal{P}(J_i,m^2_i,k_i;\tilde{J}_i,\tilde{m}^2_i,l_i)=\prod_i\frac{i\delta_{J_i,\tilde{J}_i}\delta(m^2_i+\tilde{m}^2_i)\delta_{k_i,l_i}}{C_{J_i}+m_i^2}\times
\nonumber \\ \times 
\begin{array}{c}\includegraphics[width=1.5cm]{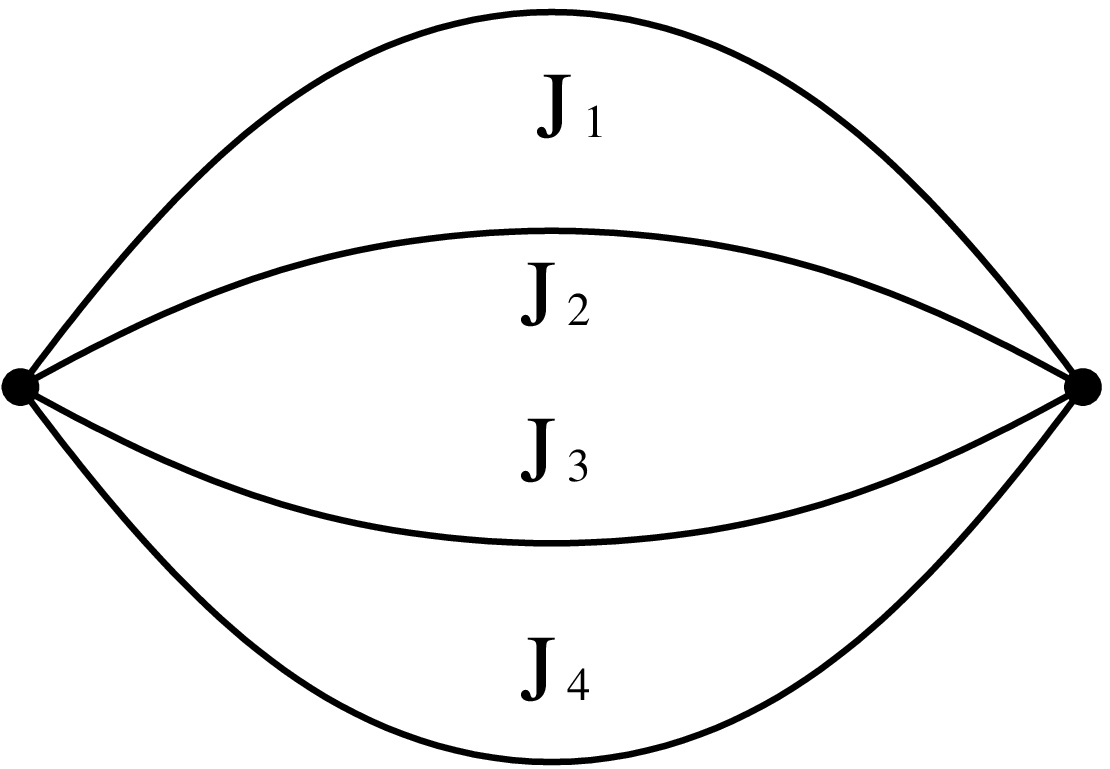}
\end{array}\delta\left(\sum_i m^2_i\right),
\ees
i.e. a product of Feynman propagators (with variable mass square, taking both positive and negative values, and
appropriate analytic continuation, using Feynman prescription)
one for each dual face intertwined in both their variables by the \lq
eye diagram\rq and by a delta function relating the masses. Had we
imposed the $P_h$ projectors also in the kinetic term the resulting propagator would have been the same, but with the inverse of the eye diagram instead of the
eye diagram in the above formula, with an extra factor $\Delta_{J_i}$
for each face. The difference in the results between the two choices
is therefore exactly the same as in the usual GFTs. 
The vertex amplitude for the various possible choices of interactions, labeled by the $\alpha_i$s and the $\mu$s, is easily read out of the action to be:
\bes
\lefteqn{\mathcal{V}\left( g_{ij},s_{ij}, \mu,\alpha_i\right)=\prod_{i=1}^{5} \int
  dg_i\int ds_i \prod_{i<j}\int_{SU(2)} dh_{ij}} \nonumber \\ &&\prod_{i<j} \left\{\theta(\alpha_i\alpha_j(s_{ij}+s_i-s_j-s_{ji}))\right.\;\;\;\;\;\;\;\;\;
\nonumber \\ &&\left. K(g_{ij}h_{ij}g_i,
g_{ji}h_{ji}g_j;\mu\alpha_i\alpha_j(s_{ij}+s_i-s_j-s_{ji}))\right\} \nonumber
\ees
in configuration space. In momentum space it reads:
\bes
\lefteqn{\mathcal{V}\left( J_{ij}, m^2_{ij},k_{ij},\{\alpha_i\}\right)
  = \prod_{i<j=1}^{5}\left( \frac{i\alpha_i\alpha_j}{\Delta_{J_{ij}}(\mu\alpha_i\alpha_j
    C_{J_{ij}}+ m^2_{ij})}\right) \times} \nonumber \\
&&\delta_{J_{ij},J_{ji}}\delta(m^2_{ij}+m^2_{ji})\delta_{k_{ij},k_{ji}}\prod_{i=1}^{5}\delta\left( \sum_{j\neq
  i=1}^{5}m^2_{ij}\right)
\begin{array}{c}\includegraphics[width=1.5cm]{4simp.eps}
\end{array} \nonumber
\ees
with the representations of $Spin(4)$ entering the formula being only the class I ones, because of the $P_h$ projectors in the action. 
The partition function of the theory is then expanded in Feynman
diagrams (we restrict for simplicity to the special case of all $\lambda$'s being equal) as:
\be
\mathcal{Z}_{gen}(\lambda)=\sum_{\Gamma_{\alpha,\mu}}\frac{\lambda^N}{sym(\Gamma_{\alpha,\mu})}Z(\Gamma_{\alpha,\mu}).   
\ee
and the new type of spin foam models for given 2-complex $\Gamma_{\mu,\alpha}$, now characterized by its orientation encoded in the $\alpha$'s and the $\mu$'s, is given by:
\bes
Z(\Gamma)=\prod_f\left( \int dm_f^2\sum_{J_f}\right)\prod_f
\Delta_{J_f} \prod_e \left[ \prod_{f\mid e} \frac{1}{C_{J_f} + m^2_f}
  A_e(J_{f})\,  \right] \nonumber \\ \prod_v \left\{ \prod_{e\mid
  v}\left[\delta\left(\sum_{f\mid e}m^2_f\right)\right]\prod_{f\mid v}
\left[\frac{1}{\mu\alpha_{1}\alpha_{2}C_{J_f} + m^2_f}\right]
\begin{array}{c}\includegraphics[width=1.5cm]{4simp.eps}
\end{array} \right\} \nonumber
\ees
where there are two independent variables for each face of $\Gamma$: a
mass variable and a  representation of $Spin(4)$. The $A_e(J_{f \mid
  e})$ is either the eye diagram function or its inverse according to
whether one imposes the $P_h$ projectors in the kinetic term of the
action or not, and we have dropped two redundant deltas relating mass
variables for each dual edge, arising from a redundant use of $P_s$
projectors in the action. We see that the above choice of action for
this generalised field produces a spin foam model with the causal vertex amplitudes of
\cite{feynman} in momentum space, as desired. The amplitude $Z_{gen}$ is in general a
complex number (recall the analytic continuation implicit in the above
expression), while the usual spin foam models are real, in agreement
with the interpretation of them as defining an analogue of the Feynman
propagator and of the Hadarmard function for (third quantized) quantum
gravity.
As usual, there is some freedom, given the type of field one is
considering and the symmetries one wants to impose, in the choice of
the action for the theory; in particular, given the
interaction term chosen above to reproduce the causal spin foam
models, one can choose different kinetic terms. A possible choice,
that becomes almost a necessity when dealing with real fields, is to
take as kinetic term the \lq square\rq of the previous one:
\be
\prod_{i=1}^{4}\int
dg_i\int ds_i
\phi^{-\mu\alpha}(g_i,s_i)\left[\prod_i\left(
  +\partial^2_{s_i}+(\nabla_{i})^2\right)\right]\phi^{\mu\alpha}(g_i,s_i) \label{kin2}
\ee
(the 4th order nature of the differential operator not being of
concern, given that we are not dealing with fields living in
spacetime, but we are at a more abstract level). This produces in
perturbation theory the same type of model as the above, but with
factors of the type $\frac{1}{C_J^2-(m^2)^2}$ instead of
$\frac{1}{C_J+m^2}$ in the edge amplitudes, the rest being
unchanged. This choice is appealing also because it includes two
opposite poles in the edge amplitude, which agrees with the structure
of the vertex amplitude.
Another interesting choice is the
so-called {\it static ultra-local} reduction of the above action
\cite{caianiello}, in which one drops every derivative in the kinetic
term, in a sense thus dropping any propagating degree of freedom, and
is left with a kinetic term given by a product of delta functions
only. This implies also dropping any correlation between mass
variables and representations in the propagator, and results in an edge
amplitude that is the same as in usual spin foam models (only
with an additional delta function cosntraining th sum of the mass
variables related to that edge). This choice of kinetic term {\it
  plus} a simple modification of the interaction term, i.e.
dropping the $\theta$ functions appearing in it (thus dropping the
causality restriction), which is equivalent
to using Hadamard propagators instead of Feynman propagators in each dual face, gives:
\bes
Z(\Gamma)=\prod_f\left( \int dm_f^2\sum_{J_f}\right)\prod_f
\Delta_{J_f} \prod_e
A_e(J_{f})\;\;\;\;\;\;\;\;\;\;\;\;\;\;\;\;\;\;\;\;\;\;\;\;\;&
\nonumber \\ \hspace{-1.5cm}\prod_v \left\{ \prod_{e\mid
  v}\delta\left(\sum_{f\mid e}m^2_f\right)\prod_{f\mid v}
\delta\left(\mu\alpha_{1}\alpha_{2}C_{J_f} + m^2_f\right)
\begin{array}{c}\includegraphics[width=1.5cm]{4simp.eps}
\end{array} \right\}  \nonumber \\ \hspace{-0.5cm}=\sum_{J_f}\prod_f
\Delta_{J_f} \prod_e A_e(J)\prod_v \left\{ \prod_{e\mid
  v}\delta\left(\sum_{f\mid
  e}\alpha^f_{e'}C_{J_f}\right)\begin{array}{c}\includegraphics[width=1.5cm]{4simp.eps}
\end{array} \right\}, \nonumber
\ees
i.e. the {\it usual} spin foam models only with two extra constraints
for each edge of $\Gamma$, one for each vertex touched by it, relating
the representations associated to the triangles of the tetrahedron
dual to it, and depending on the orientation of the other tetrahedra
in the same 4-simplex, given by the corresponding $\alpha$'s. These
extra constraints may be welcomed given that one needs extra constraints on the areas of the triangles used as
fundamental variables in simplicial gravity to have a classical theory
that is equivalent to (2nd order) Regge calculus\cite{area}. This deserves to be
explored. In any case, this shows the exact sense in which the usual
spin foam models, thus the type of quantum gravity transition
amplitudes they correspond to, can be obtained in the generalised
group field theory formalism we are presenting. Notice that the same
amplitudes, modulo additional delta divergences, could be obtained
using the kinetic term \Ref{kin2} but using Hadamard propagators for
each face in the edge instead of Feynman ones, i.e. going \lq
on-shell\rq, while dropping at the same time the $\theta$ functions in
the vertex term; this would be a somewhat awkward construction and not
easy to interpret in terms of usual field theoretic perturbation
theory (perturbative expression of the anticommutator of field
operators?), but it may nevertheless shed more light on the nature of
the usual spin foam models from the point of view of this generalised
formalism.
Notice also that dropping the $\theta$ functions in the vertex term
turns each vertex amplitude into a real function
and the poles of the complex functions appearing in it into
zeros of delta functions, i.e. means going \lq on-shell\rq, which in
turns has the effect of confining the orientation dependence of the
amplitude in the additional constraints on the face representations.
As we have discussed, one of the motivations for developing this
generalised GFT formalism was to obtain models that 
could be more easily related to classical simplicial actions. We can
indeed check that the vertex amplitude we have proposed above has an expression very
close to the exponential of the Regge action in 1st order formalism
\cite{causal,feynman}. Going to a mixed representation in terms of the
variables $g_i$ and $m_i^2$, i.e. performing the Fourier transform
with respect to the proper time variables, this looks like:
\bes
\prod_{f_{ij}\mid v} \int_{0}^{+\infty}ds_f\,\theta(\alpha_i\alpha_j
s_f)\, K(\vartheta_{f_{ij}},\mu_v\alpha_{i}\alpha_{j}s_f)e^{-im_f^2s_f}=
\nonumber \\ =\prod_{f_{ij}\mid v}
\frac{1}{4\pi}\frac{1}{\sin\vartheta_{f_{ij}}}e^{i\mu_v
  \sqrt{1-\mu\alpha_{i}\alpha_{j}m^2_{f_{ij}}}\vartheta_{f_{ij}}}.
\ees
Interpreting the quantity
$\sqrt{1-\mu\alpha_{i}\alpha_{j}m^2_{f_{ij}}}$ as the area of the
triangle dual to the face $f_{ij}$, which is consistent with the
stationary point analysis of usual spin foam models (since on-shell,
i.e. when $C_{J_{f_{ij}}} = -\mu \alpha_i\alpha_j m^2_{f_{ij}}$, the
above quantity is equal to $2J_{f{ij}}+1$), we would have then a
vertex amplitude given by the exponential of the Regge action in 1st
order form \cite{causal,feynman}, {\it if} the
angle $\vartheta_{f_{ij}}$ was the dihedral angle between the tetrahedra
$i$ and $j$. However, this angle measures the holonomy around the portion of the
face dual to their shared triangle inside the 4-simplex $v$, so it
receives a contribution not only from the group elements labelling the
parallel transport around dual edges {\it inside} the 4-simplex (that
indeed correspond to the dihedral angle), but also from the boundary
holomies represented by the group elements being the arguments of the
field in configuration space. Therefore we clearly get the exponential
of the Regge action for each 4-simplex in the special case of flat
boundary holonomies, but more work is needed to show that this is
also the result in the general case and after integrating out the
boundary variables to obtain the quantum amplitude for
the whole Feynman graph, i.e. for the whole simplicial complex. The
above calculation however suggests that this is likely to happen.

\section{Conclusions}

We have
constructed a generalised formalism for group field theories that
extends the usual one by the use of a field depending on a set of
extra proper time variables and their conjugate mass variables. The
first motivation was the wish to be
able to derive both a-causal and causal transition amplitudes for
quantum gravity, of the type presented in \cite{feynman}. We have
shown that indeed one can derive this type of spin foam models with an
appropriate choice of interaction term in the generalised group field
action. We have also identified two different choices for the kinetic
term that can be used in conjunction with this interaction term to
define causal transition amplitudes (spin foam models), and producing
slightly different edge amplitudes. The static-ultra-local limit of
the generalised group field theory action, together with a removal of
the causal restriction on the proper time variables in the interaction
term, reproduces naturally the usual (a-causal) spin foam models, thus
clarifying their role and interpretation within the generalised
formalism. Finally, we have shown how the vertex amplitudes for the
generalised group field theory with causality restrictions are given
by complex functions of the variables of the theory, with an
expression very closely related to the exponential of the Regge action
for simplicial gravity, thus hinting to an expression of this type for
the complete Feynman graph amplitudes. If this is verified, maybe for
slightly different choices of the group field theory action, the
nature of the models will be manifestly that of a
simplicial third quantization of gravity and the generalised formalism
will be naturally understood as a unified framework for as different
approaches to non-perturbative quantum gravity as loop quantum
gravity, spin foam models, quantum Regge calculus and dynamical
triangulations, and one would be left to study in detail how each of
them can be obtained and understood within the generalised formalism.

Several lines for future research are opened by these results. We list
some of them. One should investigate
other possible actions within the
generalised formalism, and then study
symmetries and amplitudes of the resulting spin foam models. 
The geometric meaning of the extra variables (proper time and
mass) should be clarified in
various ways: by analysing the relation with simplicial actions for
gravity, extablishing the link between the quantum amplitudes of the
generalised GFT and the exponential of the Regge action, and then
understanding the meaning of the alternative expressions that the
generalised formalism provides for the same quantitites (a \lq
parametrised Regge action\rq ?); by studying parametrised formulations
of gravity and the consequent extension of superspace
\cite{greensite}; by considering alternative descriptions and
quantizations of the \lq\lq quantum tetrahedron\rq\rq \cite{bb}.
Also, the form of the action in this generalised
formalism may facilitate the analysis of the classical equations of
motion as well as the Hamiltonian analysis of the theory. The coupling
of particles in the 3d version of this generalised models, along the
lines of \cite{particle}, may produce Feynman diagrams for the matter fields
that involve correctly feynan propagators for the field, instead of
Hadamard ones as in \cite{particle}. Finally, it will be crucial to
show how other approaches to quantum gravity can be obtained from this generalised GFT formalism.


\begin{thebibliography}{99}
\bibitem{review} D. Oriti, Rept.Prog.Phys. 64, 1489 (2001),gr-qc/0106091;
\bibitem{alex} A.Perez,Class.Quant.Grav. 20,R43(2003),gr-qc/0301113;
\bibitem{mikecarlo} M. Reisenberger, C. Rovelli,
  Class. Quant. Grav. 18, 121 (2001),gr-qc/0002095 
\bibitem{BC} J. W. Barrett, L. Crane, Class.Quant.Grav. 17 (2000)
  3101-3118, gr-qc/9904025;
\bibitem{laurentkirill} L. Freidel, K. Krasnov, J.Math.Phys. 41 (2000)
  1681-1690, hep-th/9903192;
\bibitem{DP-F-K-R} R. De Pietri, L. Freidel, K. Krasnov, C. Rovelli,
  Nucl.Phys. B574 (2000) 785-806, hep-th/9907154;
\bibitem{causal} E. Livine, D. Oriti, Nucl. Phys. B 663, 231 (2003), gr-qc/0210064;
\bibitem{feynman} D. Oriti, Phys. Rev. Lett. 94, 111301 (2005), gr-qc/0410134;
\bibitem{teitelboim} C. Teitelboim, Phys. Rev. D 25, 3159 (1982);
\bibitem{halliwellhartle} J. J. Halliwell, J. B. Hartle, Phys. Rev. D
  43, 1170 (1991);
\bibitem{giddingsstrominger} S. Giddings, A. Strominger,
  Nucl.Phys. B321, 481 (1989);
\bibitem{asymp} J. W. Barrett, R. M. Williams,
  Adv. Theor. Math. Phys. 3, 209-215 (1999), gr-qc/9809032;
\bibitem{DT} J. Ambjorn, J. Jurkiewicz, R. Loll, Phys. Rev. D 72,
  064014 (2005), hep-th/0505154;
\bibitem{laurentgft} L. Freidel, hep-th/0505016;
\bibitem{iogft} D. Oriti, in {\it Mathematical and Physical Aspects of
  Quantum Gravity},J.Tolksdorf, B.Fauser eds, Birkhauser (2006);
\bibitem{Feynman} R. P. Feynman, Phys. Rev. 80, 440-457 (1950);
\bibitem{hostler} L. Hostler, J. Math. Phys. 21, 2461 (1980);
\bibitem{DP-P} R. De Pietri, C. Petronio, J. Math. Phys. 41, 6671-6688
  (2000), gr-qc/0004045;
\bibitem{P-R} A. Perez, C. Rovelli, Nucl. Phys. B 599, 255-282 (2001),
  gr-qc/0006107;
\bibitem{caianiello} E. R. Caianiello, G. Scarpetta,  Nuovo Cim. A 22, 448(1974)
\bibitem{area} J. Makela, R. M. Williams, Class. Quant. Grav. 18, L 43
  (2001), gr-qc/0011006;
\bibitem{greensite} A. Carlini, J. Greensite,
  Phys. Rev. D 55, 3514-3524 (1997), gr-qc/9610020. 
\bibitem{bb} J. Baez, J. W. Barrett, Adv. Theor. Math. Phys. 3, 815-850 (1999), gr-qc/9903060
\bibitem{particle} L. Freidel, D. Louapre, Class. Quant. Grav. 21,
  5685 (2004), hep-th/0401076; L. Freidel, D. Oriti, J. Ryan,
  gr-qc/0506067; D. Oriti, J. Ryan, in preparation.
\end{thebibliography}
\end{document}